\documentclass[10pt,letterpaper]{article}
\usepackage{opex3}
\usepackage{epsfig}
\usepackage{amsmath}
\usepackage{amssymb}
\usepackage{mathrsfs}

\begin{document}

\title{Photonic crystal fiber with a hybrid honeycomb cladding}

\author{Niels Asger Mortensen}
\address{Crystal Fibre A/S, Blokken 84, DK-3460 Birker\o d, Denmark}
\email{asger@mailaps.org}

\author{Martin Dybendal Nielsen}
\address{Crystal Fibre A/S, Blokken 84, DK-3460 Birker\o d, Denmark\\COM, Technical University of Denmark, DK-2800 Kongens Lyngby, Denmark}

\author{Jacob Riis Folkenberg, Christian Jakobsen, and Harald R. Simonsen}
\address{Crystal Fibre A/S, Blokken 84, DK-3460 Birker\o d, Denmark}
\homepage{http://www.crystal-fibre.com}


\begin{abstract}
We consider an air-silica honeycomb lattice and demonstrate a new approach to the formation of a core defect. Typically, a high or low-index core is formed by adding a high-index region or an additional air-hole (or other low-index material) to the lattice, but here we discuss how a core defect can be formed by manipulating the cladding region rather than the core region itself. Germanium-doping of the honeycomb lattice has recently been suggested for the formation of a photonic band-gap guiding silica-core and here we experimentally demonstrate how an index-guiding silica-core can be formed by fluorine-doping of the honeycomb lattice.
\end{abstract}





\begin{figure}[b!]
\begin{center}
\epsfig{file=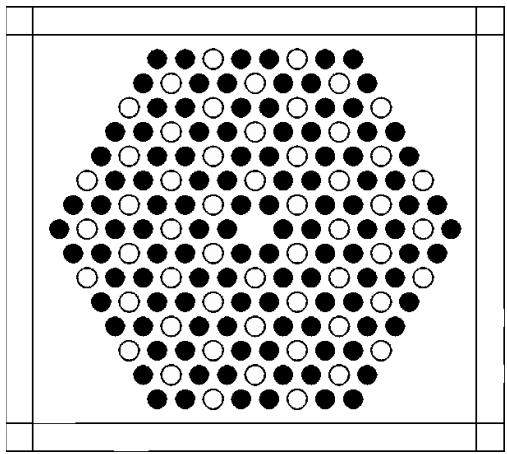, width=0.5\textwidth,clip}
\end{center}
\caption{Cross-section of the PCF with air-holes indicated by filled circles and the fluorine doped regions indicated by open circles. The perfectly-matched layers employed in finite-element simulations are also indicated. } 
\label{fig1}
\end{figure}

\section{Introduction}

In their basic form photonic crystal fibers (PCF) typically consist of fused silica with an arrangement of air-holes running along the full length of the fiber (for a recent review we refer the reader to Ref.~\cite{knight2003} and references therein). Traditionally, triangular~\cite{knight1996} or honeycomb~\cite{knight1998} cladding arrangements of the air-holes have been considered with the core defect formed by removing or adding an additional air-hole in the lattice, respectively. This is of course the most obvious way to form a defect in a regular lattice. However, for the honeycomb lattice (see Fig.~1) there is at least one alternative approach which involves additional use of index-altering dopants. Recently, L\ae gsgaard and Bjarklev~\cite{laegsgaard2003d} suggested how a low-index band-gap guiding core could be formed in a 
germanium doped honeycomb lattice by absence of doping in the core region. Here, we suggest doping by fluorine which results in an index-guiding core. Fluorine-doped PCFs have previously been considered by Mangan {\it et al.}~\cite{mangan2001} who fabricated a triangular air-hole cladding PCF with a fluorine-doped silica core region. At sufficiently short wavelengths the core index is lower than the effective index of the cladding and the PCF is anti-guiding, but as the wavelength is increased the effective index of the cladding decreases and eventually becomes lower than the core index so that light is guided in the core region. In the present work we use fluorine doping to form a novel large-mode area PCF and the proposed fiber structure may be an alternative to existing large-mode area designs employing a triangular air-hole arrangement in the cladding \cite{knight1998_LMA,mortensen2003a,nielsen2003d}.

\begin{figure}[t!]
\begin{center}
\epsfig{file=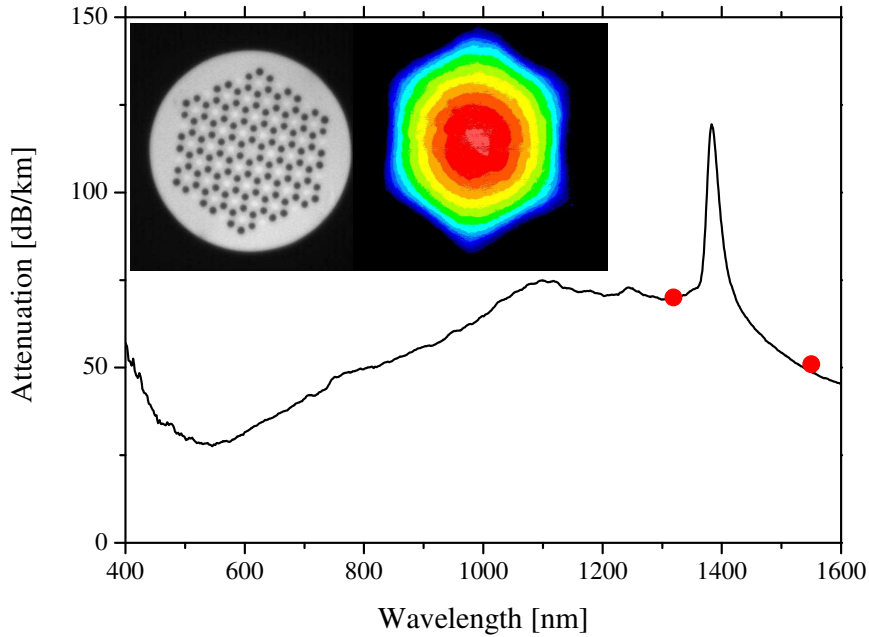, width=0.9\textwidth,clip}
\end{center}
\caption{Spectral loss measured by a standard white-light cut-back technique. OTDR measurements at $\lambda= 1319\ {\rm nm}$ and 1550 nm are also indicated by red dots. The measurements are performed with 200 m of fiber on a spool with a radius of 8 cm. The left insets show an optical micrograph of the fiber end-facet with the dark circular regions showing the air holes and the light regions showing the fluorine-doping in the silica background. The right inset shows a near-field image of the fundamental mode at $\lambda= 635\ {\rm nm}$.} 
\label{fig2}
\end{figure}

\section{Fiber design and fabrication}

We consider the structure in Fig.~\ref{fig1} where fluorine doped regions (of diameter $d_f$) are incorporated in a honeycomb lattice of air-holes (of diameter $d$ and nearest-neighbor spacing $\Lambda$). The core region is formed by the absence of doping in the central region of the structure. At sufficiently short wavelengths the cladding states will avoid the fluorine-doped regions and the effective cladding index will in some sense resemble that for a triangular arrangement of holes whereas at longer wavelengths the field averages over the fluorine-doped and pure silica regions so that the effective index resembles that of a honeycomb lattice (with a slightly down-shifted background index). The defect region has six-fold rotational symmetry and thus supports a doubly degenerate fundamental mode \cite{steel2001}.

We have fabricated the proposed fiber by the stack-and-pull method \cite{knight1996} with hexagonal stacking of fluorine-doped rods (step-index like doping profile) and fused silica tubes and rods. The inset in Fig.~\ref{fig2} shows an optical micrograph of a typical fiber cross section. The fiber has an outer diameter of $175\,{\rm \mu m}$ in order to reduce micro-bending deformations at short wavelengths \cite{nielsen2003e} and it is coated with a standard single layer acrylate coating. The fluorine-doped regions are of diameter  $d_f/\Lambda \sim 0.7$ with an index $n_f$ suppressed by $\delta n \sim  5 \times 10^{-3}$ relative to the index $n_s$ of silica. The pitch is $\Lambda\simeq 10.1\,{\rm \mu m}$ and the relative air-hole diameter is $d/\Lambda=0.64$.

\section{Fiber characterization}

To facilitate coupling of light to the core region a SMF28 was spliced to 200 m of the PCF. The spectral attenuation was measured by the cut-back technique using a white-light source and an optical spectrum analyzer. Fig.~\ref{fig2} shows the spectral attenuation of the PCF. Data from OTDR measurements at $\lambda = 1319\,{\rm nm}$ and 1550 nm are also included and good agreement with the cut-back data is observed. Cut-back transmission experiments on a short length of PCF have revealed no signatures of high-order modes.

\begin{figure}[t!]
\begin{center}
\epsfig{file=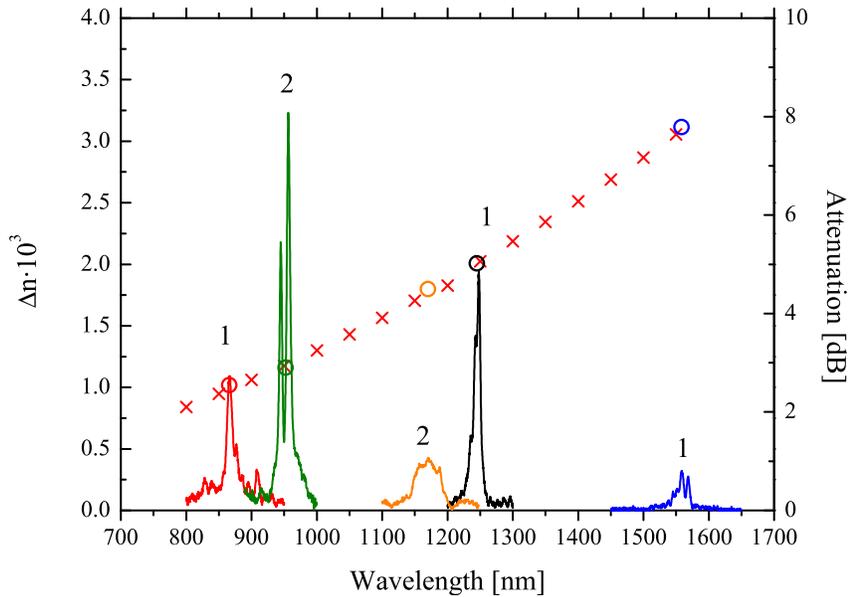, width=0.9\textwidth,clip}
\end{center}
\caption{Mode-spacing (left axis) derived from periodic micro-deformation spectra (right axis). Red crosses indicate values of $\Delta n$ from numerical simulations while the solid curves are the measured attenuation peaks induced by periodic micro-deformations. The number above each peak indicate if the peak is of 1st or 2nd order and the open circles represent the corresponding mode spacing calculated from the measurements.} 
\label{fig3}
\end{figure}

The fiber has a low loss regime at short wavelengths and a regime with high loss above 1000 nm extending to the O-H absorption peak. In order to further analyze the modal properties we have studied the mode-spacing which can be derived from micro-deformation measurements. For a detailed description of the method and the interpretation of experimental data we refer the reader to Ref.~\cite{nielsen2003} and references therein. Fig.~\ref{fig3} shows loss spectra (right axis) for various periodic micro-deformations and the derived mode-spacings, $\Delta n$ are also included (left axis). The figure also includes numerical data calculated with the finite-element method incorporating perfectly matched layers \cite{saitoh2002}, see Fig.~\ref{fig1}. In agreement with the experimental observations, our simulations also suggest that the PCF is broad-band single mode \cite{birks1997} in the sense that high-order modes have a negligible spacing ($\ll 10^{-4}$) to cladding modes. 

In order to understand the spacing between guided modes and cladding modes we apply first-order perturbation theory to the guided modes. Treating the absence of fluorine-doping in the core as a perturbation, $\delta\varepsilon = n_s^2- n_f^2\simeq 2n_s\delta n$, we may estimate the shift $\Delta n$ in mode-index with respect to the cladding modes. From standard perturbation theory (see {\it e.g.} Ref.~\cite{johnson2002}) we get

\begin{equation}
\Delta n = \frac{c}{2v_g}\frac{\big< {\boldsymbol E}\big|\delta\varepsilon\big|   {\boldsymbol E}\big> }{\big< {\boldsymbol E}\big|\varepsilon\big|   {\boldsymbol E}\big> } \simeq \frac{cn_s}{v_g}\frac{\big< {\boldsymbol E}\big|\delta n\big|   {\boldsymbol E}\big> }{\big< {\boldsymbol E}\big|\varepsilon\big|   {\boldsymbol E}\big> }
\end{equation}
where $c$ is the velocity of light, ${\boldsymbol E}$ is the unperturbed electrical field, and $v_g$ is the group velocity. For a high-order mode the field-intensity is strongly suppressed at the center of the core region \cite{kuhlmey2002,mortensen2003c,folkenberg2003,mortensen2004b} and since $d_f/\Lambda$ is not too large the mode has a very small overlap with the region where fluorine is absent. This results in a negligible increase in effective index $\Delta n$ with respect to the cladding modes. In other words, localization in the core is no big energetic advantage for the high-order modes. For the fundamental mode the situation is opposite since it has a large field-intensity at the center of the core and indeed we find a mode-spacing $\Delta n$ comparable to $\delta n$. 

The mode-spacing picture in Fig.~\ref{fig3} suggests that the overall loss has little relation to bending-induced scattering loss (we have also verified this by changing the bending radius $R$) and since confinement loss can be excluded (we have verified this numerically) it is likely that the overall high background originates from various contamination added during the stack-and-pull fabrication process. We believe that the background loss level can be lowered similarly to the recent achievements in triangular cladding large-mode area PCFs \cite{nielsen2003d}.

\section{Conclusion}

We have studied a new approach to the formation of a core defect in the honeycomb structure by fluorine-doped regions added to the cladding. At sufficiently short wavelengths the cladding states will avoid the fluorine-doped regions and the effective cladding index will in some sense resemble that for a triangular arrangement and light will be confined to the core region where fluorine-doping is absent. 

We believe that hybrid cladding designs could be an interesting direction which allows a higher degree of modal/dispersion engineering and the particular structure studied in this paper could also be interesting for photonic band-gap fiber devices employing liquid crystals \cite{larsen2003}.

\section*{Acknowledgments}

N.~A. Mortensen thanks J. L\ae gsgaard (COM, Technical University of Denmark) for stimulating discussions and L. Gregersen (Comsol A/S) for technical support. M.~D. Nielsen acknowledges financial support by the Danish Academy of Technical Sciences.

\end{document}